# VAN KAMPEN MODES FOR BUNCH LONGITUDINAL MOTION

A. Burov, FNAL[*], Batavia, IL 60510, U.S.A.

*Abstract*

Conditions for existence, uniqueness and stability of bunch steady states are considered. For the existence uniqueness problem, simple algebraic equations are derived, showing the result both for the action and Hamiltonian domain distributions. For the stability problem, van Kampen theory is used [1-3]. Emerging of discrete van Kampen modes show either loss of Landau damping, or instability. This method can be applied for an arbitrary impedance, RF shape and beam distribution function Available areas on intensity-emittance plane are shown for resistive wall wake and single harmonic, bunch shortening and bunch lengthening RF configurations.

## MAIN EQUATIONS

Let $H(z,p)$ be a Hamiltonian for longitudinal motion inside RF bucket distorted by the wake field:

$$H(z,p) = \frac{p^2}{2} + U(z) + V(z,t);$$
$$U(z) = U_{\mathrm{rf}}(z) - \int \lambda(z')W(z-z')dz'; \quad (1)$$
$$V(z,t) = -\int \rho(z',t)W(z-z')dz'.$$

Here $z$ and $p$ are the offset and the momentum of a particle, $U(z)$ is the steady state potential with $U_{\mathrm{rf}}(z)$ as its RF part, $\lambda(z)$ is steady state linear density, $W(z)$ is the wake function, $V(z,t)$ and $\rho(z,t)$ are small perturbations of the potential well and linear density. For the potential well $U(z)$, action $I$ and phase $\varphi$ variables can be found:

$$I(H) = \frac{1}{\pi}\int_{z_{\min}}^{z_{\max}} \sqrt{2(H-U(z))}dz;$$
$$\Omega(I) = \frac{dH}{dI}; \quad \frac{dz}{d\varphi} = \frac{\sqrt{2(H-U(z))}}{\Omega(I)}. \quad (2)$$

The linear density $\lambda$ and its perturbation $\rho$ can be related to steady state phase space density $F(I)$ and its perturbation $f(I,\varphi,t)$:

$$\lambda(z) = \int F(I)dp;$$
$$\rho(z,t) = \int f(I,\varphi,t)dp. \quad (3)$$

Below, the steady state distribution $F(I)$ is treated as an input function, determined either by cooling-diffusion kinetics, or by injection. The perturbation $f(I,\varphi,t)$ satisfies the Jeans-Vlasov equation [4]:

$$\frac{\partial f}{\partial t} + \Omega(I)\frac{\partial f}{\partial \varphi} - \frac{\partial V}{\partial \varphi}F'(I) = 0. \quad (4)$$

Set of equations (1-4) assumes given input functions $U_{\mathrm{rf}}(z)$, $W(z)$ and $F(I)$, while the steady state solution $U(z)$, $I(H)$, $\lambda(z)$ and all the eigenfunctions of the dynamic Jeans-Vlasov equation (4) are to be found.

To obtain the steady state solution, the following set of three equations is to be solved:

$$U(z) = U_{\mathrm{rf}}(z) - \int \lambda(z')W(z-z')dz' \equiv U_{RHS}[\lambda];$$
$$I(H) = \frac{1}{\pi}\int_{z_{\min}}^{z_{\max}} \sqrt{2(H-U(z))}dz \equiv I_{RHS}[U]; \quad (5)$$
$$\lambda(z) = 2\int_{U(z)}^{H_{\max}} \frac{F(I(H))}{\sqrt{2(H-U(z))}}dH \equiv \lambda_{RHS}[I,U].$$

For any given input functions $U_{\mathrm{rf}}(z)$, $W(z)$ and $F(I)$, the solution can be numerically found by means of a quasi-time method. Indeed, let it be assumed that initially there is no wake, so that the entire potential well is equal to the RF potential $U(z)=U_0(z)=U_{\mathrm{rf}}(z)$. With that, initial action and linear density functions $I_0(H)$ and $\lambda_0(z)$ can be found from the 2nd and 3rd equations of the set (5). Then the following iteration procedure can be applied:

$$U_n(z) = U_{n-1}(z) - \varepsilon\left(U_{n-1}(z) - U_{RHS}[\lambda_{n-1}]\right);$$
$$I_n(H) = I_{RHS}[U_n]; \quad (6)$$
$$\lambda_n(z) = \lambda_{RHS}[I_n, U_n]; \quad n = 1, 2, \ldots$$

If the solution exists, the process converges to it provided the convergence parameter $\varepsilon>0$ is sufficiently small. When the steady state is found, the dynamical stability analysis can be performed by means of Jeans-Vlasov equation (4). Following Oide and Yokoya [5], the eigenfunctions may be expanded in Fourier series over the synchrotron phase $\varphi$:

$$f(I,\varphi,t) = e^{-i\omega t}\sum_{m=1}^{\infty}\left[f_m(I)\cos m\varphi + g_m(I)\sin m\varphi\right]. \quad (7)$$

With the zero-phase at the left stopping point,

$$z(I,\varphi=0) = z_{\min}(I); \quad z(I,\varphi=\pi) = z_{\max}(I);$$
$$z(I,-\varphi) = z(I,\varphi); \quad p(I,-\varphi) = -p(I,\varphi), \quad (8)$$

this yields an equation for the amplitudes $f_m(I)$:

$$\left[\omega^2 - m^2\Omega^2(I)\right]f_m(I) = -2m^2\Omega(I)F'(I)\sum_{n=1}^{\infty}\int dI' V_{mn}(I,I')f_n(I'); \quad (9)$$

$$V_{mn}(I,I') = -\frac{2}{\pi}\int_0^\pi d\varphi \int_0^\pi d\varphi' \cos(m\varphi)\cos(n\varphi')W(z(I,\varphi) - z(I',\varphi')).$$

The matrix elements $V_{mn}(I,I')$ can be also expressed in terms of the impedance $Z(q)$. After [6]

$$W(z) = -i\int_{-\infty}^{\infty}\frac{dq}{2\pi}\frac{Z(q)}{q}\exp(iqz) \quad (10)$$

it results in

$$V_{mn}(I,I') = -2\operatorname{Im}\int_0^\infty dq \frac{Z(q)}{q}G_m(q,I)G_n^*(q,I');$$

$$G_m(q,I) \equiv \int_0^\pi \frac{d\varphi}{\pi}\cos(m\varphi)\exp[iqz(I,\varphi)]. \quad (11)$$

Note that there is no bunch-to-bunch interaction in the formulas above; this long-range wakes are omitted for simplicity sake. Equations (9-11) reduce the integro-differential Jeans equation (4) to a standard eigen-system problem of linear algebra after the action integral in Eq.(9) is approximately substituted by a sum.

## STEADY STATE SOLUTION

The algorithm of Eq. (6) allows finding numerical solution of the steady state problem. In this section, a problem of existence and uniqueness of that solution is considered.

It is well-known that below a certain temperature threshold, there is no thermodynamic equilibrium (no solution of the Haissinski equation [7]) for the space charge wake above transition, $W(z) \sim \delta(z)$ [6,8]. For this case though, the distribution function is given in the Hamiltonian domain, $F = C\exp(-H/T)$. Thus, its normalization constant $C$ is yet to be found from the normalization condition, $2\pi C\int_0^\infty \exp(-H(I)/T)dI = 1$, which is to be added to the entire set of equations and has to be solved jointly with them. If the temperature $T$ is low enough, the normalization condition leads to an algebraic equation having no solutions. The same problem would appear for any distribution function expressed in terms of the Hamiltonian, not the action. For the Hoffman-Pedersen distribution $F = C\sqrt{H_{\max} - H}$ this is shown in Ref. [6]. For the space charge above transition, and some other wakes, the bunch momentum spread and average Hamiltonian turn out to be limited from below: for given RF and intensity they cannot be smaller certain value for any longitudinal emittance. That is why it may be wrong to assume arbitrary distribution function in terms of Hamiltonian. On the contrary, equations (5) assume distribution density as a function of action, so it is a priory explicitly defined and normalized, and limitations for Hamiltonian-domain functions may not apply for this case.

A following estimation allows to see when Eqs. (5) do have a solution. Let $\bar{I}$ be rms bunch emittance, and $l$ be the rms bunch length; then, the rms momentum spread is estimated as $\bar{p} \cong \bar{I}/l$, and the average synchrotron frequency $\Omega \cong \bar{p}/l \cong \bar{I}/l^2$. From another side, $\Omega^2 - \Omega_0^2 \propto \operatorname{Im}(Z(l^{-1}))/l^2$, where $\Omega_0$ is the bare RF synchrotron frequency. Combination of these two expressions yields (compare with Ref. [6], p. 285):

$$\bar{I}^2 q^4 = 1 + q^2 \operatorname{Im} Z(q);$$

where $q=1/l$ is the inverse bunch length to be found from this equation, $q$ is measured in inverse radians of RF phase. Emittance $\bar{I}$ is dimensionless, its value in conventional eV·s units can be found after multiplication by a factor of $E_0\Omega_0/(\eta\omega_{\mathrm{rf}}^2)$, where $E_0 = \gamma mc^2$ is the beam energy. The dimensionless impedance $Z(q)$ of this paper, Eqs. (1,10), relates to the conventional $Z_\parallel(q)$ of Ref. [6] as $Z(q) = DZ_\parallel(q)$ with the intensity factor $D = Nr_0\eta c\omega_{\mathrm{rf}}^2/(\Omega_0^2\gamma C)$, where $N$ is the bunch population, $r_0$ - the classical radius, $\eta = \gamma_t^{-2} - \gamma^{-2}$ – the slippage factor, $\omega_{\mathrm{rf}}$ – RF angular frequency, $\gamma$ – relativistic factor, $C$ – the machine circumference. Note that this equation does not pretend to give an exact solution for the bunch length; it is an estimation showing existence of the solution and its dependence on the parameters. It follows that the solution exists if the wake singularity is not too high: at high frequencies the impedance may not grow too fast, $\lim_{q\to\infty}\operatorname{Im}Z(q)/q^2 = 0$, what is true for all known cases. For non-monotonic impedances, in general there may be several solutions though. For the space charge and the resistive wall impedances there is always a unique steady state.

There is an obvious reason though, why Eqs. (5) may still have no solution: the bucket has a limited acceptance; thus, it cannot hold a bunch which emittance is higher than that. Moreover, in many cases, wake fields reduce bucket capacity. This could lead to some beam loss to DC, even if the bunch would fit within the bare RF bucket.

In case the distribution function is given as a function of Hamiltonian, with its average value $\bar{H}$ the steady state estimation writes as

$$\bar{H}q^2 = 1 + q^2 \operatorname{Im} Z(q).$$

For that case, the solution existence is not intensity-limited only for slow-growing or bunch-lengthening impedances, when $D \lim_{q \to \infty} \operatorname{Im} Z(q) \leq 0$; what is not satisfied for the space charge above transition and the resistive wall below transition. For these impedances, there are either no solutions, or two of them. For the last case, the two solutions have the same Hamiltonian distribution but different phase space density.

## VAN KAMPEN MODES

More than half a century ago, N. G. van Kampen found eigen-system of Jeans-Vlasov equation for infinite plasma [1-3]. The spectrum was found consisting of continuous and discrete parts. Continuous spectrum essentially describes single-particle motion, with addition of a proper plasma response. Frequency band of the continuous spectrum is one of the incoherent frequencies; for any velocity $v$ within the distribution function, there is a continuous van Kampen mode with a frequency $kv$, where $k$ is the wave number. Continuous modes are described by singular functions in the velocity space, underlying their primary relation to single-particle motion. In these terms, Landau damping results from phase mixing of van Kampen modes of the continuous spectrum. On the contrary to the continuous spectrum, the discrete one not necessarily exists; there may be no discrete modes. If the discrete spectrum is not empty, all its modes are described by regular functions, and some of the modes do not decay. Indeed, since the original equations (analogue of Eq. (9)) are with real coefficients, the mode frequencies are either real or forming complex-conjugate pairs. The first case corresponds to a loss of Landau damping (loss of LD, or LLD), while the second describes an instability. Plasma with monotonic distribution density was shown to be always stable; discrete modes of LLD type may only appear if the distribution function is of a finite width. The instability is possible for non-monotonic distributions only.

Most of the plasma results are applicable to bunches. However, two issues make a difference of bunches from plasma. First, beam particle interaction may be described by various wake functions, it is not reduced to pure Coulomb forces. And second, the frequency spectrum for bunch particles is always limited, while in plasma the velocity spectrum may be considered as infinite, at least formally.

Eigen-modes of Jeans-Vlasov equation for bunch longitudinal motion were first considered by A. N. Lebedev [9]. Although the suggested formalism was not numerically tractable, an important result was analytically obtained: it was proved that for the space charge impedance above transition, a bunch steady state is always stable. Numerically tractable algorithm was suggested more than twenty years later by Oide and Yokoya [5].

For parabolic RF potential, van Kampen modes were analyzed for power wake [5], capacitive [10], broad-band wakes [5,10] and modified inductive wake [11]. For that RF, rigid bunch oscillations at the unperturbed synchrotron frequency is always a solution of equation of motion [10]. Indeed, single-particle equations of motion can be written as

$$\ddot{z}_i + \Omega_0^2 z_i = \sum_j W'(z_i - z_j); \quad i, j = 1, ..., N.$$

The solution can be presented as a sum of a steady-state-related part $\hat{z}_i$ and a small perturbation $\tilde{z}_i$. It is clear that with the rigid-bunch motion $\tilde{z}_i = A\cos(\Omega_0 t)$ that equation is satisfied. Typically there is no incoherent frequency of that value, since all of them are either suppressed or elevated by the potential well distortion; thus, this mode is discrete. It is not necessarily though: as it was shown in Ref [10], for broad band impedance model, core and tail incoherent frequencies may go with intensity in opposite directions, so the rigid-bunch mode may be covered by incoherent frequencies, proving to be Landau-damped. In Ref. [12], loss of Landau damping was analyzed assuming it is the rigid-bunch mode which is losing its Landau damping; similar assumption was done in Ref. [10]. As it is shown in the next section, generally it is not correct: action dependence of the emerging discrete mode can be very different from the rigid-bunch mode; it is always so when the RF frequency spread is taken into account.

Without interaction, there are no discrete modes for Eqs.(9), all of them belong to continuous, or incoherent, spectrum, $\omega = m\Omega(I)$. If the bunch intensity is low enough, weak head-tail approximation may be applied, omitting terms with different azimuthal numbers. In this case, it is straightforward to show that for monotonic distributions, $dF/dI<0$, and for symmetric potential wells, $U(-z)=U(z)$, Eq. (9) reduces to one with a symmetric matrix; thus, in this case, all its eigenvalues are real. Since there are no unstable modes, all the discrete modes, if any, belong to LLD type. In practice there are always some losses, and so the distorted potential well $U(z)$ is always somewhat asymmetric. However, my attempts to find weak head-tail instability in numerical solutions for resistive wake and purposely asymmetric RF potentials never succeeded. This contradicts to results of Ref [13], where radial mode coupling was found within a simplified 2-mode analysis. To save CPU time, I so far limited my stability analysis by the weak head-tail approximation and the dipole azimuthal mode $m=1$.

This paper takes into account two possible reasons for beam intensity limitations: reduction of bucket acceptance by wake fields and loss of Landau damping (LLD).

## RESULTS FOR RESISTIVE WALL

In this section, intensity limitations for resistive wall impedance are summarized. The energy is supposed to be above transition; otherwise, the space charge impedance normally cannot be neglected. RF potential is supposed to be as

$$U_{\rm rf}(z) = (1 - \cos z) + \alpha_2(1 - \cos 2z)/4. \quad (12)$$

Three options for the second RF harmonic are considered: single harmonic (SH), $\alpha_2=0$, bunch shortening (BS), $\alpha_2=1$, and bunch lengthening (BL), $\alpha_2=-1$. Here $\Omega_0$ is zero-amplitude synchrotron frequency provided by the first harmonic only. For SH case, the RF bucket acceptance (maximal action) in dimensionless units is $8/\pi \approx 2.54$. The energy offset is related to the dimensionless momentum as $\delta E / E_0 = -p\Omega_0 / (\eta \omega_{rf})$, time offset is $z/\omega_{rf}$. In these dimensionless units, the wake function and impedance of a round chamber with radius $b$ and conductivity $\sigma$ are presented as [6]:

$$W(s) = -k / \sqrt{-s} \ ;$$

$$Z(q) = k(1 - i\,\mathrm{sgn}\,q)\sqrt{\pi |q|/2} \ ;$$

$$k = \frac{Nr_0 \eta \omega_{rf}^2}{\pi \gamma b \Omega_0^2} \sqrt{\frac{\omega_{rf}}{\sigma}} \ .$$

An example with the parabolic potential shows that wake fields act more on incoherent frequencies than on the coherent ones. For the parabolic potential, the first discrete mode does not depend on the impedance at all. Thus, at certain threshold, a first discrete mode jumps out of the continuous spectrum, since its frequency is not suppressed or increased as much as the incoherent frequencies are. For the SH and BS RF, above transition, lowest-amplitude particles are mostly excited for this mode, since their frequencies are closer to the coherent mode. That is why an assumption of rigid-bunch discrete mode is not generally correct.

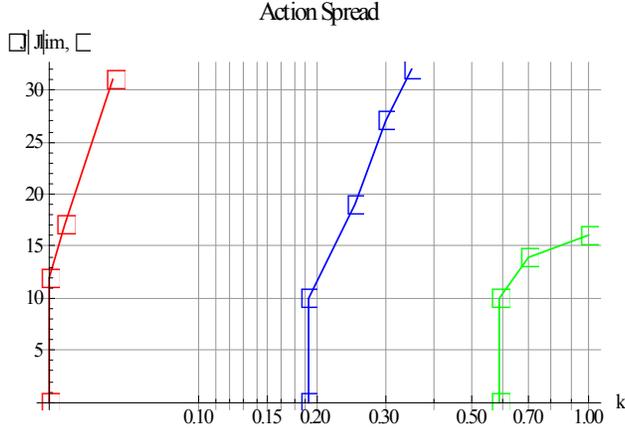

Figure 1: Relative width of the discrete mode $\sigma_I/I_{lim}$, %, versus intensity parameter $k$ for distribution $F(I) \sim (I_{lim} - I)^{1/2}$ and emittances $I_{lim} = 0.5$, 1.0 and 1.5 (red, blue and green).

In Fig.1, relative width of that discrete mode $\sigma_I/I_{lim}$ is shown, where

$$\sigma_I = \sqrt{\frac{\int dI f^2(I)(I - \bar{I})^2}{\int dI f^2(I)}}, \quad \bar{I} \equiv \frac{\int dI f^2(I) I}{\int dI f^2(I)},$$

as a function of the intensity parameter $k$ for the Hoffman-Pedersen distribution $F(I) \sim (I_{lim} - I)^{1/2}$ and three values of the emittance $I_{lim}$. It is clear that after the threshold the mode widens rather fast, what indicates its weak sensitivity on details of the distribution as soon as the bunch emittance is given.

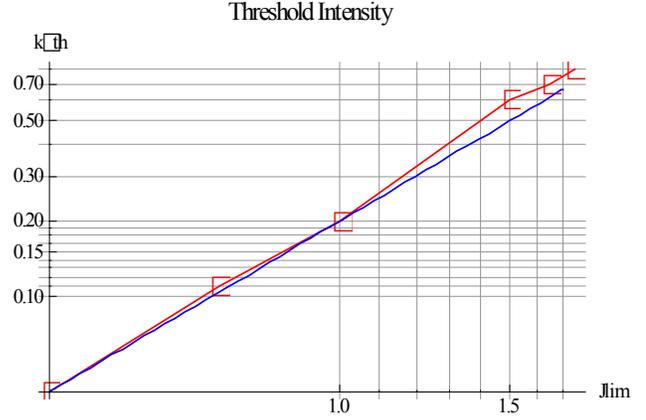

Figure 2: Threshold intensity parameter $k_{th}$ for SH RF versus the bunch emittance for the same distribution function as in Fig. 1.

LLD threshold for the intensity parameter $k$ versus the emittance $I_{lim}$ is presented in Fig.2. The blue fitting line: $k_{th} = 0.2 I_{lim}^{9/4}$. Although the emergent discrete mode is far from being similar to the rigid-bunch motion $\propto \sqrt{I/(I_{lim} - I)}$, its power 9/4 agrees with the simplistic rigid-bunch model [14]. For BS RF bucket, the discrete mode looks rather similar to the SH case; the LLD threshold is nicely fitted as $k_{th} = 0.55 I_{lim}^{5/2}$.

Contrary to SH and BS RF cases, for the BL RF incoherent frequency is not monotonic function of action, it has a maximum at $I = I_m \approx 1.5$. That is why, for the considered case of effectively repulsive wake, the discrete mode emerges from the tail particle frequencies, if the bunch limiting emittance is smaller than that, $I_{lim} < I_m$. For BL RF, emergence of the discrete mode is sensitive to the tails of the distribution: even a tiny tail covering the coherent frequency returns Landau damping, so it kills that discrete mode. If the bunch emittance is not that small, $I > I_m$, the discrete mode emerges above the incoherent maximum. Since this mode emerges outside the entire bucket area of the incoherent frequencies, Landau damping cannot be returned by tiny perturbations of the distribution function. That is why this kind of LLD, which cannot be cured by tiny corrections of the distribution function, is called here as radical LLD. To avoid that tail ambiguity, only radical LLD is taken as a real stability limit.

Figure 3 shows radical LLD limitations for BL RF, for two different distributions. Note that LLD limits the available acceptance by $I_m \approx 1.5$, while the entire BL bucket area is about twice higher. Stability limitation associated with the maximum of incoherent synchrotron frequencies was first pointed out in Ref. [15]; for BL RF it was analytically found with the rigid-beam model in Ref. [12], and experimentally studied at CERN SPS (see Ref. [16] and references therein).

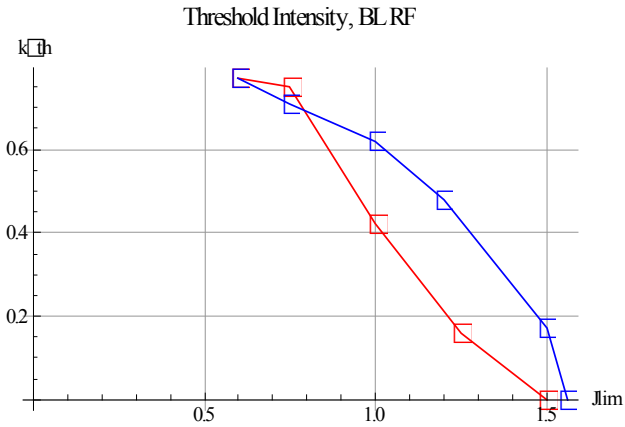

Figure 3: Threshold intensity $k_{th}$ versus emittance $I_{lim}$ for BL RF and two distribution functions: $F(I)\sim(I_{lim}-I)^{1/2}$ (red) and $F(I)\sim(I_{lim}-I)^2$ (blue).

On the $k$-$I_{lim}$ area, the availability is limited by LLD and bucket capacity. For the three RF configurations, SH, BS and BL, their areas of availability are shown in Fig. 4. The left slope of the BL (green) line is preliminary and may significantly change when azimuthally mode coupling will be taken into account. It is seen that every RF configuration has its own beneficiary area: hot and low-intensity beams better fit into SH, cold high-intensity ones are more suitable for BL, and intermediate case is for BS RF.

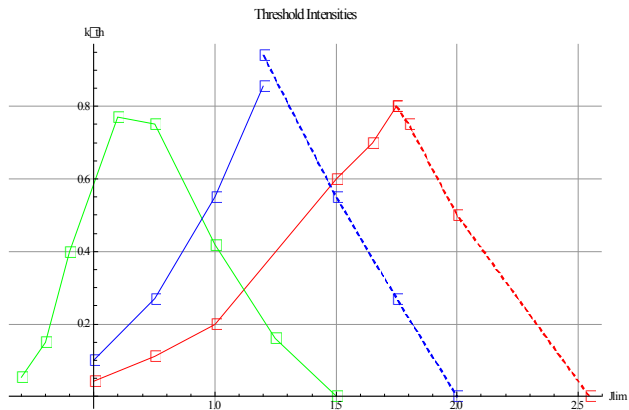

Figure 4: Intensity-emittance $k$-$I_{lim}$ areas of availability for $F(I)\sim(I_{lim}-I)^{1/2}$. Red lines are for SH, blue – for BS, green – for BL. Solid lines show radical LLD or instability, dashed – limiting bucket capacity.

## CONCLUSIONS

Language of van Kampen modes is a powerful tool for studying beam stability. Its unique efficiency reveals itself in those complicated cases, when the dielectric function cannot be obtained, as it is for the longitudinal bunch motion. Emergence of a discrete mode means either loss of Landau damping or instability. By definition, the discrete modes lie outside the continuous incoherent spectrum, but they still may stay within the bucket. In the last case, the discrete mode would disappear after a tiny portion of resonant particles would be added. However, if the discrete mode lie outside the bucket, the Landau damping cannot be restored by tiny perturbation of the particle distribution; LLD is called radical in that case.

For a given bunch emittance and RF voltage, the intensity is limited either by reduction of the bucket acceptance or by (radical) LLD. In this paper, results are presented for longitudinal bunch stability in weak head-tail approximation and resistive wall impedance; three RF configurations are studied: single harmonic, bunch shortening and bunch lengthening. It is shown that every RF configuration may be preferable, depending on the bunch emittance and intensity.